\documentclass{aa}  

\usepackage[]{amsmath}      
\usepackage{graphicx}
\usepackage{color}
\usepackage{txfonts}
\usepackage{url}
\usepackage{natbib}         
\bibpunct{(}{)}{;}{a}{}{,}  
\newcommand{\vc}[1]{\mbox{\boldmath{$#1$}}}  

\begin{document} 

   \title{
   Formation of planetary systems by pebble accretion and migration
   }
   \subtitle{ 
    How the radial pebble flux determines a \\ terrestrial-planet or super-Earth growth mode
    }
   
   \author{
   Michiel Lambrechts\inst{1,2} \and
   Alessandro Morbidelli\inst{2} \and
   Seth A. Jacobson\inst{3} \and
   Anders Johansen\inst{1} \and \\
   Bertram Bitsch\inst{4} \and
   Andre Izidoro\inst{5} \and
   Sean N. Raymond\inst{6} 
   }

   \institute{
	   Lund Observatory, Department of Astronomy and Theoretical
	   Physics, Lund University, Box 43, 22100 Lund, Sweden\\
	   \email{michiel@astro.lu.se}
           \and
           Laboratoire Lagrange, UMR7293, Universit\'e C\^ote d'Azur, CNRS,
           Observatoire de la C\^ote d'Azur, Boulevard de l'Observatoire, 06304
           Nice Cedex 4, France
           \and 
           Department of Earth and Planetary Sciences, Northwestern University,            Technological Institute, F293/4, 
           2145 Sheridan Road, Evanston, IL 60208-3130, USA
           \and
           Max-Planck-Institut f\"ur Astronomie, 
           K\"onigstuhl 17, 69117 Heidelberg, Germany
           \and
           UNESP, 
           Univ. Estadual Paulista - Grupo de Din\`amica Orbital Planetologia, 
           Guaratinguet\`a, CEP 12.516-410, S\~ao Paulo, Brazil
           \and
           Laboratoire d'Astrophysique de Bordeaux, 
           CNRS and Universit\'e de Bordeaux, 
           All\'ee Geoffroy St. Hilaire, 33165 Pessac, France
           }

   \date{}

\abstract{
Super-Earths -- planets with sizes between the Earth and Neptune -- are found
in tighter orbits than the Earth's around more than one third of main sequence
stars. 
It has been proposed that super-Earths are scaled-up terrestrial planets that
also formed similarly, through mutual accretion of planetary embryos, but in
discs much denser than the solar protoplanetary disc.
We argue instead that terrestrial planets and super-Earths have two clearly
distinct formation pathways that are regulated by 
the disc's pebble reservoir.
Through numerical integrations, which combine pebble accretion and N-body
gravity between embryos, we show that a difference of a factor of two in the
pebble mass-flux is enough to change the evolution from the terrestrial to the
super-Earth growth mode.
If the pebble mass-flux is small, then the initial embryos within the ice line
grow slowly and do not migrate substantially, resulting in  a widely spaced
population of $\sim$\,Mars-mass embryos when the gas disc dissipates. 
Then, without gas being present, the embryos become unstable due to
mutual gravitational interactions and a small number of terrestrial planets are
formed by mutual collisions.
The final terrestrial planets are at most $5$ Earth masses.
Instead, if the pebble mass-flux is high, then the initial embryos within the
ice line rapidly become sufficiently massive to migrate through the
gas disc. 
Embryos concentrate at the inner edge of the disc and growth accelerates
through mutual merging. 
This leads to the formation of a system of closely spaced super-Earths in
the $5$ to $20$ Earth-mass range, bounded by the pebble isolation mass.
Generally, instabilities of these super-Earth systems after the disappearance
of the gas disc trigger additional merging events and dislodge the system from
resonant chains.
Thus, the key difference between the two growth modes is whether embryos grow
fast enough to undergo significant migration.
The terrestrial growth mode produces small rocky planets on wider orbits like those in
the Solar System whereas the super-Earth growth mode produces planets in
short-period orbits inside 1\,AU, with masses larger than the Earth, that
should be surrounded by a primordial H/He atmosphere, unless subsequently lost
by stellar irradiation.
The pebble flux -- which controls the transition between the two growth modes
-- may be regulated by the initial reservoir of solids in the disc or the
presence of more distant giant planets that can halt the radial flow of
pebbles.
}

   \keywords{
	     Planets and satellites: formation -- 
	     Planets and satellites: dynamical evolution and stability -- 
	     Planets and satellites: composition -- 
	     Planets and satellites: terrestrial planets -- 
             Protoplanetary disks
             }

   \maketitle
%

\section{Introduction}

Super-Earths are, broadly speaking, exoplanets with masses, or radii,
intermediate to those of the Earth and Neptune. 
The occurrence rate of such exoplanets is high: more than $30$\% of sun-like
stars harbour super-Earth planets within $100$-day orbits \citep{Mayor_2011,
Petigura_2013, Zhu_2018}. 
Occurrence rates are even higher, by approximately a factor 3, around
lower-mass M-dwarf stars \citep{Mulders_2015}.
Systems of multiple super-Earths are common and typically have low
eccentricities \citep[$e< 0.05$, ][]{Xie_2016} and low mutual inclinations
\citep[$i \lesssim$ $10^{\circ}$,][]{Lissauer_2011b,Johansen_2012,Zhu_2018}. 

The composition of these super-Earths are observationally difficult to
determine, but their mass budget appears to be dominated by a rocky interior.
From the subset of well-characterized planets, it is inferred that planets with
radii below $1.8$ Earth radii (R$_{\rm E}$) are mainly rocky in
composition, based on planetary structure models
\citep{Rogers_2015,Lopez_2014}.
Larger planets -- above $2$\,R$_{\rm E}$ and mass of about $5$ Earth mass
($M_{\rm E}$) -- are consistent with having primordial H/He envelopes that make up $1$\,\% and $20$ of the total mass \citep{Hadden_2017}.
The composition of the core of these planets with gas envelopes is not well known, but can be probed around close-in planets that likely lost their envelope through irradiation from the host star.  
Models of envelope loss favor rocky interiors to explain the lack of planets with radii between $2$ to $4$\,R$_{\rm E}$ on highly irradiated orbits \citep{Lundkvist_2016}. 
Similarly, the lack of planets with radii around $1.8$\,R$_{\rm E}$ within $100$-day orbits \citep{Fulton_2017} may be best explained when envelope loss occurs around cores with a rocky, as opposed to water-rich, composition \citep{Owen_2017, Jin_2018}.

It is not obvious that super-Earths, even when rock-dominated, could have
formed in a way similar to the Earth.
This is because the Earth is characterized not only by its rocky composition
of $67.5$\% silicates and $32.5$\% iron, but also by its slow formation.
The growth of the Earth likely took place over a timescale of several tens of
Myr, based on the age constraints on the Moon-forming impact
\citep{Touboul_2007,Kleine_2009,Jacobson_2014, Barboni_2017}. 
In contrast, Mars formed within $3$ to $5$\,Myr, according to radiogenic dating 
\citep{Nimmo_2007,Dauphas_2011}.
Thus, the formation timescale of the Earth greatly exceeds the average gas-rich
phase of protoplanetary discs of $3$ to $5$\,Myr \citep{Haisch_2001}, while
Mars could have formed within the gas phase.
Terrestrial planet formation beyond the mass of Mars is therefore believed to
have taken largely place in a gas-free environment, where the Earth is the
product of mutual collisions of planetary embryos which were roughly Mars-sized
at the time the gas disc dissipated (see \citealt{Morbidelli_2012b}
and \citealt{Raymond_2014} for a review).
The last of these collisions corresponds then to the Moon-forming event \citep{Hartmann_1975,Cameron_1976}.

This gas-free growth-mode of the Earth from Mars-sized embryos had several implications for its final properties. 
For instance, Mars-mass protoplanets do not migrate significantly in the
proto-planetary disc \citep{Tanaka_2002}, which explains why the Earth could
remain relatively far from the Sun. 
Similarly, Mars-mass protoplanets cannot capture substantial H and He envelopes
directly from the gas disc \citep{Mizuno_1978} and such tenuous envelopes erode
easily during the subsequent series of impacts \citep{Schlichting_2015}. 
This explains why the Earth does not have a primitive atmosphere, but instead
one outgassed from its interior, dominated by much heavier gases than hydrogen
\citep{Schaefer_2010}.

Super-Earths must have experienced a different, more rapid, growth process. 
Their larger rocky cores argue for an increased mass reservoir resulting in
faster embryo growth.  
This leads to larger embryos before disc dissipation, which
necessarily introduces significant inward migration \citep{Ogihara_2015}. 
In turn, the resulting concentration of embryos can speed up further growth by collisions.
Also, these larger embryos can capture significant primitive H/He
atmospheres, like those inferred around large super-Earths.

The goal of this paper is to develop a unified model for the formation of, on
the one hand, Earth-like planets with 
temperate orbits, and on the other hand, close-in super-Earths. 
In order to do so, we will consider embryo growth that is mainly driven by the
accretion of inwards-drifting pebbles.
The accretion cross section of an embryo for pebbles that feel gas drag can
greatly exceed the cross section for gravitationally focused planetesimals
\citep{Ormel_2010, Lambrechts_2012}. 
Therefore, core growth by pebble accretion from the radial mass flux of pebbles
that settle to the midplane and drift inward through the disc can exceed
classical planetesimal accretion rates \citep{Lambrechts_2014b, Levison_2015a, Levison_2015b, Lin_2018}.
Specifically, in this work, we will only consider embryos that are located
within the ice line. 
This also implies that the pebbles are ice-free. 
Thus, the focus is on the growth of rocky embryos.

We find that the critical parameter dividing Earth-like formation from
migration-assisted formation of super-Earths is the integrated pebble-mass flux
through the inner protoplanetary disc. 
The available mass in pebbles depends on many parameters, the initial total
disc mass, the initial dust-to-gas ratio, the radial extent of the disc, and
also the possible presence of planets larger than approximately $10$\,M$_{\rm
E}$ that block the flow of pebbles to the inner disc \citep{Morby_2012,
Lambrechts_2014a}.
In this work, we will assume the integrated pebble-mass flux to be a unique free
parameter, for simplicity.

This paper is structured as follows. 
Section 2 describes the set-up of the simulations and explains how we take into
account the presence of the gas disc, planetary migration, and pebble
accretion. 
Validation tests and a more detailed description of the pebble accretion formulae can be found in appendix\,\ref{app:num}.

Section 3 shows that the divergent evolution of a system of growing embryos
depends on the mass carried by the integrated pebble-flux. 
We find that a low pebble-flux leads to the slow formation of small planetary embryos that do not migrate significantly in the disc. 
At most these embryos grow to about $3$ Mars-masses. 
The increase in the pebble-flux, by less than a factor of 2 with respect to
this case, drastically bifurcates the evolution of the system. 
As expected, the embryos grow faster and become more massive as they start to migrate towards the star. 
This migration-assisted growth-mode leads to planets of several Earth masses
near the inner edge of the disc within the lifetime of the gas disc. 

Section 4 follows the systems in their evolution after the removal of the gas disc.
All the extended systems of numerous, small planetary embryos become unstable
and lead to the formation of Earth-like planets on a timescale of tens of Myr,
with a sequence of giant impacts analogous to that characterizing the formation
of our planet. 
We find that the most massive Earth-like planets generated in this way are between $2$ and $5$\,M$_{\rm E}$. 
For the super-Earth systems we find that
they can undergo a dynamical instability shortly after
disc dissipation, typically within $10$\,Myr, 
similar to
\citet{Terquem_2007},
\citet{Ogihara_2009},
\citet{Ida_2010},
\citet{Cossou_2014},
\citet{Izidoro_2017}, 
\citet{Carrera_2018} and
\citet{Ogihara_2018a}.

This leads to the reduction of the final number of planets,
a few merging events, and the acquisition of non-resonant orbits with mutual
spacings that are more consistent with observations.
We note however that the fraction of our super-Earth systems that become unstable after gas removal is much larger than in \citet{Izidoro_2017} who find only half of the super-Earth chains to be become unstable. 
We find instead $>90$\% unstable cases, which appears to be in better agreement
with the observations. This is due to our super-Earth systems forming more
compact during the gas disc phase, 
due to the combined effects of migration and pebble accretion, the latter of which \citet{Izidoro_2017} neglects.

Wrapping up these results, we argue in Section 5 for a differentiation between
Earth-like planets and super-Earths, not based on a simplistic mass-threshold
or difference in bulk composition, but instead based on the growth history of
the planet.
Because the growth history of a given body cannot be observed, we suggest a
number of combined observational criteria to distinguish between these two
categories of planets: the mass, the orbital architecture and the presence of a
primitive atmosphere, if the planet is not strongly irradiated by the host
star. 
In Section 6 we discuss the available mass reservoir of pebbles in the inner disc and summarize the assumptions made in this work.
We conclude with our main findings in Section 7.

This paper comes as part of a set of three papers on the formation of planets by
N-body simulations that take pebble accretion and planetary migration into
account.
The other two papers differ from this work in that they also consider embryo
growth outside of the ice line.
\citet{TRILOGY_ANDRE} show that the inclusion of icy embryos leads to
super-Earth systems that can reproduce quantitatively the observed orbital
distribution of Kepler systems. 
However the predominantly icy composition of these planets is in apparent
contrast with the inferred rocky composition of Kepler planets
\citep{Owen_2017, Jin_2018}.
Finally, \citet{TRILOGY_BERT} discuss the case where the pebble mass-flux is
large enough such that some of the icy embryos can turn into giant planets in
wide orbits, as occurred in the Solar System.
It builds on the work by \citet{Levison_2015a} and \citep{Bitsch_2015b}, but
uses a self-consistent modeling for the growth of both giant planets and
super-Earths.
This work shows that the migration of giant planets into the region interior to
$1$\,AU can be prevented if embryos form sufficiently far from the ice line,
outside $30$\,AU.
Embryos closer to the ice line, between $5$ and $10$\,AU, migrate into the
inner disc, unless Type-II migration rates are reduced compared to nominal
values, as expected in discs with comparable mass but lower viscosity
\citep{Kanagawa_2018,Robert_2018}.

Taken together, this trilogy of papers should provide a quite
comprehensive view of planet formation and evolution revealing a broad spectrum
of possibilities. 
They have in common that pebble accretion is the main process fueling the
initial growth of proto-planets and that the formation of the final planetary
systems is the result of a complex interplay between mass growth and dynamical
evolution.

\section{Methods}

\subsection{N-body code}

We have used a modified version of the N-body code {\texttt SyMBA}, which uses
a symplectic algorithm that allows adaptive time-steps for close encounters
\citep{Duncan_1998,Levison_2012}.
Collisions are modelled as events that always lead to perfect merging.
We have added prescriptions to the N-body code for the protoplanetary gas disc, 
planet-disc interaction and the presence and accretion of pebbles. 
We describe these in turn below.

\subsection{Disc model}

Here, we use a simple model to describe the gaseous component of the
protoplanetary disc.
The aspect ratio of the gaseous disc, which is equivalent to the ratio of the sound speed $c_{\rm s}$ to Keplerian velocity $r\Omega_{K}$, is given by
\begin{align}
  H/r = 0.04 \, .
\end{align}
We thus have a flat aspect ratio with orbital distance (zero flaring), which is approximately realised in the inner disc, where viscous heating dominates over irradiation \citep{Bitsch_2015a,Ida_2016}.
The gas surface density is given by 
\begin{align}
  \Sigma_{\rm g} = 
   610\,\left(\frac{r}{\rm AU} \right)^{-1/2} 
   \times \exp \left[-\frac{t}{t_{\rm disc}} \right] \, {\rm g/cm}^{2} \,.
\end{align}
The slope of the surface density is chosen such that the disc has a constant viscously-driven gas accretion rate through the disc with 
$\dot M_{\rm gas} = 3\pi \Sigma_{\rm g} \nu$, 
assuming a constant $\alpha$ value for the viscosity $\nu = \alpha c_{\rm s}^2/\Omega_{\rm K}$. 
Our disc mass is thus less centrally concentrated that in the more crude disc
estimate based on the Minimum Mass Solar Nebula, which has
$\Sigma_{\rm g} \approx 1700 (r/{\rm AU})^{-3/2}\,{\rm g/cm}^{2}$
\citep{Weidenschilling_1977, Hayashi_1981}.
We consider here a low-viscosity disc with $\alpha = 10^{-4}$. 
If one would only consider viscously-driven gas accretion, the initial
accretion rate would be $\dot M_{\rm gas}=  7 \times 10^{-10}$\,M$_{\odot}$/yr,
which would be in lower range of observed gas accretion rates around young
stars \citep{Manara_2016}.
However, current magnetohydrodynamical disc modelling efforts argue that gas
accretion is mainly wind-driven, regulated by active layers above the midplane.
This support the use of low values of $\alpha$ for the midplane turbulence, without increasing the surface density of the disk $\Sigma$ as $1/\alpha$
\citep{Suzuki_2009,Bai_2014,Lesur_2014,Gressel_2015,Bai_2017}. 
This modern view of disks  appears to be observationally supported by
measurements of particle settling \citep{Pinte_2016}.
Because our aim here is to focus on the dynamic interplay between migration and
accretion, we have not used complex time-dependent disc models
\citep{Bitsch_2015a, Suzuki_2016, Ogihara_2018a,Ogihara_2018b}.
Instead, we mimic the dissipation of the gas disc with a simple exponential
decay of the gas surface density on a time scale of $t_{\rm disc} =1$\,Myr and we consider the gas disc phase to last $3$\,Myr in our simulations \citep{Haisch_2001}.
Importantly, note that we assume that the temperature in the disk does not decrease with time. 
Therefore, we do not consider here a drifting snowline that can cross the growing protoplanets \citep[see however companion papers][]{TRILOGY_ANDRE,TRILOGY_BERT}.

\subsection{Type-I migration and e/i-damping}

In order to model planet-disc interaction, we make use of prescriptions that model how gas affects embryos in the disc.
Because we do not consider planets that grow much beyond $\approx
10$\,M$_{\rm E}$ we limit ourselves to consider type-I migration as well as
inclination and eccentricity damping.
We use the force formulation approach by \citet{Paps_2000},
\begin{align}
  \vc a_{\rm tidal} = - \frac{\vc v}{t_{\rm m}} - 2 \frac{\vc v \cdot \vc r}{r^2 t_{e} } - 2 \frac{(\vc v \cdot \vc k ) \vc k}{t_i} \,.
\end{align}
Here, $\vc k$ is the unit vector in the vertical direction, $\vc r, \vc v$ and
$ \vc a$ are the radial position, velocity and acceleration.
The first term represents the migration, with $t_{m}$ the migration timescale.
The second term damps the eccentricity on a timescale $t_{\rm e}$ and the third term corresponds to inclination damping on a timescale $t_{\rm i}$.
All three timescales are proportional to the wave-damping timescale \citep{Tanaka_2002}, 
\begin{align}
  t_{\rm wave} = 
  \frac{M_\odot}{M_{\rm p}} 
  \frac{M_{\odot}}{\Sigma_{\rm g} a_{\rm p}^2} 
  \left(\frac{H}{r}\right)^4 \Omega_{\rm p}^{-1} \,,
\end{align}
but are modified in a complex fashion by their dependency on the eccentricity and inclination of the body \citep{Bitsch_2010,Cossou_2013,Fendyke_2014}. 
Here, $M_{\rm p}$ and $M_{\odot}$ are, respectively, the planet and star mass.
The e,i-dependent formulation for these timescales were taken from fits to hydrodynamical simulations by \citet{Creswell_2008} 
\footnote{
We found it important to use the formulation by \citet{Creswell_2008}, because it also covers the high eccentricity and inclination ($e>h/r$) cases. 
Additionally, we also noted that the eccentricity and inclination damping timescales fits of \citet{Creswell_2008} are best used joined with the \citet{Paps_2000} force approach. Applying the  \citet{Creswell_2008} damping formula to the type-I force formulation by \citet{Tanaka_2002} and \citet{Tanaka_2004} can lead to unexpected results for high e,i cases.
}.
The migration timescale we used includes the contribution of the differential
Lindblad torque and the co-rotation torque, under the approximation of an
isothermal disc \citep{Tanaka_2002}.
Because we use a simple power-law disc model, there are no corotation torque traps and migration is always directed inwards.
More precisely, for our choice of disc model, the wave-damping timescale
($t_{\rm wave}$) and migration timescale ($t_{\rm m} \propto (H/r)^{-2} t_{\rm
wave}$) do not depend on the orbital radius $r$. 
This implies embryos of equal mass would migrate in sync, until they approach the
inner disc edge.
Companion papers \citet{TRILOGY_ANDRE} and \citet{TRILOGY_BERT} include the entropy-related corotation torque and consider more complex disc models that have trapping regions.
Future work could also explore in more detail the role of 
embryo heating \citep{Benitez_2015}, 
dynamical corotation torques \citep{Paardekooper_2014,McNally_2017,Fung_2018,McNally_2018} 
and surrounding pebbles \citep{Benitez_2018} onto the migration rates of
embryos.

To mimic the edge of the gas disc, we reduce migration rates close to a trap radius of $r_{\rm trap} = 0.1$\,AU. 
In practice, when embryos approach within $2\times r_{\rm trap}$ the migration rate is reduced by a factor
\begin{align}
  c_{\rm red} = \sin \left( \frac{\pi}{2} \frac{r-r_{\rm trap}}{r_{\rm trap}} \right) \,.
\end{align}
In this way, without reducing the eccentricity and inclination damping rates,
we smoothly bring the migration of the embryos to a halt.
We do not aim to  model the complex (non-ideal) magnetohydrodynamics that sculpts the inner edge of the disc \citep{Romanova_2006,Flock_2017}.
Another trapping radius may be related to the transition to the inner MRI
active region \citep{Chatterjee_2014}.
Thus our choice of $r_{\rm trap}$ is somewhat arbitrary and therefore our final
results in terms of semi-major axis distribution can be, crudely, rescaled by
$r_{\rm trap}/(0.1\,{\rm AU})$.
Moreover, the disc edge likely moves outwards as the disc evolves, further
complicating the picture \citep{Liu_2017}.

Finally, we note that we do not stop the drift of pebbles at the trap radius,
there is thus no pebble pile-up. 
Pebbles keep drifting until they reach the edge of the simulation where they
are lost to the sun.

\subsection{Embryo distribution}
To avoid the numerical cost of simulating too many bodies, we start simulations with Moon-mass embryos ($M_{\rm embr,0} = 0.01$\,M$_{\rm E}$). 
In this way we also avoid the complications related to where, how, and with which size distribution planetesimals form \citep{Johansen_2015,Simon_2016} and how the first Moon-mass embryos emerge from these planetesimal seeds. 
Planetesimals have to be sufficiently massive, exceeding approximately the mass
of the dwarf planet Ceres, to be efficient in accreting pebbles
\citep{Lambrechts_2012,Visser_2016}. Therefore, planetesimal collisions may aid
in driving initial embryo growth \citep{Johansen_2015,JL2017}.
In total, we start out with $25$ Embryos, which thus represents in total $0.25$\,M$_{\rm E}$. 

We distribute the embryos radially from $0.5$ to $3$\,AU, in a logarithmic fashion. 
Because the ratio between neighbouring orbital radii $a_{i+1}/a_{i}$ is
constant, the initial embryo surface density is relatively steep ($\Sigma_{\rm
emb} \propto r^{-2}$).
The outer boundary is chosen pragmatically such that the initial embryos represent bodies that are not icy. 
In the Solar System this edge approximately corresponds to the position of the asteroid belt.
The location of the inner boundary of the embryo population was based on the
assumption that the initial embryo seeds emerged early in the disc lifetime
outside of the silicate sublimation front, which could have been as far out as
$0.5$\,AU away from the host star, when gas accretion rates onto the star were
of the order of $10^{-7}$\,M$_\odot$/yr \citep{Morby_2016}. 
This choice of the inner edge is also similar to the pragmatically chosen inner edge in terrestrial planet simulations \citep{Hansen_2009}.
Thereafter, gas accretion rates diminish and the silicate sublimation line
moves towards the inner disc edge \citep{Bitsch_2015a}. For simplicity, as our
disc has no temperature evolution, we assume that the silicate sublimation
front is sufficiently close so that we can ignore the sublimation of pebbles.


\subsection{Pebble accretion}

Pebbles are not modelled as individual N-body particle-tracers, as done in some works \citep{Kretke_2014,Levison_2015a,Levison_2015b}, because this is numerically expensive.
Instead the pebble surface density is calculated as a background field, based on
the given gas disc and pebble accretion flux.
We then calculate, for each body, how much of the passing pebbles are accreted.
A similar approach, can be found in \cite{Coleman_2017} and
\citet{Matsumura_2017}, but here we present a more detailed pebble accretion
model.

\paragraph{Pebble flux}
In this work we use a prescription for the global flux of pebbles as function of time described by
\begin{align}
  F_{\rm peb} = F_{\rm peb,0} \times  \exp \left[-\frac{t}{t_{\rm peb}} \right]\,.
\end{align}
We choose to set the decay timescale of the pebble flux equal to the disc dissipation time scale $t_{\rm peb} = t_{\rm disc}$. 
This is inspired by the observed high occurrence rate of pebbles in discs in
the $3$ to $5$\,Myr age range \citep{Ansdell_2017}.
For our nominal mass flux, we set $F_{\rm peb,nom} = 120$ \,M$_{\rm E}$/Myr,
which is on the order of the expected pebble fluxes in discs
\citep{Lambrechts_2014b}. 
However, we expect that the flux of pebbles into the inner disc may change significantly from one protoplanetary disc to the next.
For one, the available dust mass in solids may vary depending on the initial
disc mass and initial dust-to-gas ratio. 
Moreover, the evolution of the pebble flux may change depending the radial extent of the dust and the sticking efficiency of colliding particles  \citep{Brauer_2008}. 
Additionally, pebbles are likely reprocessed around ice lines, where the
volatile species sublimate \citep{Ros_2013,Morby_2015,Schoonenberg_2017}. 
Also the presence of giant planets in the outer disc, not directly modelled
in this work, can reduce, or even completely halt, the pebble flux. 
Therefore, we will simply consider the pebble flux in to the inner disc to be a
free parameter and explore different values of $F_{\rm peb,0}$ across different
simulations.
We present results from $4$ different suites of nominally 10 simulations. 
\begin{enumerate}
\item Suite ({\texttt{runf1}}) has 
$F_{\rm peb,0} =(1/3) \times F_{\rm peb,nom} = 40$\,M$_{\rm E}$/Myr, 
or a time-integrated pebble flux of $38$\,M$_{\rm E}$.
\item Suite ({\texttt{runf3}}) has 
$F_{\rm peb,0} = F_{\rm peb,nom} = 120$\,M$_{\rm E}$/Myr, 
or an integrated pebble flux of $114$\,M$_{\rm E}$.
\item Suite ({\texttt{runf5}}) has 
$F_{\rm peb,0} = (5/3) \times F_{\rm peb,nom} = 200$\,M$_{\rm E}$/Myr, 
or an integrated pebble flux of $190$\,M$_{\rm E}$.
\item And finally suite 4 ({\texttt{runf9}}) which has 
$F_{\rm peb,0} = 3 \times F_{\rm peb,nom}=360$\,M$_{\rm E}$/Myr, 
or an integrated pebble flux of $340$\,M$_{\rm E}$.
\end{enumerate}
We use the suffix ``\texttt{-1}'' to identify run number 1, and use the suffix
``\texttt{C}'' to indicate the continuation of the run after disc dissipation.

To facilitate the interpretation of our results, we also choose to simulate a
single particle population, characterized by a unique Stokes number. 
The stokes number is a non-dimensional number which expresses gas drag friction timescale with respect to the orbital period, 
\begin{align}
  \tau_{\rm f} 
   = \frac{\sqrt{2\pi} R \rho_{\bullet}}{\Sigma_{\rm g}}
\end{align}
Here, we have considered the relevant Epstein drag regime with $\rho_{\bullet}$ and $R$ corresponding to, respectively, the solid density and particle
radius of the pebble.
We consider a constant Stokes number of $\tau_{\rm f} = 3 \times 10^{-3}$. 
This is inspired by the small sizes of chondrules which make up a large mass
fraction of primitive meteorites \citep{Johansen_2015}. 
Such a constant Stokes number with orbital radius also appears to be a better
crude approximation, compared to particles with a fixed radius, to numerical
simulations of pebble growth and drift, where pebble sizes are limited by
either drift or fragmentation \citep{Brauer_2008}. 
Similarly, when considering the balance between particle growth and
drift/fragmentation in evolving discs, the Stokes number of the dominating
particles tend to only weakly change as the gas density decreases with time
\citep{Birnstiel_2012,Lambrechts_2014b}.
Therefore, we also keep the Stokes number constant in time, for simplicity.
Alternatively, one should consider a global pebble growth and evolution
model, like in \citet{Ormel_2017}, \citet{TRILOGY_ANDRE} and \citet{TRILOGY_BERT}, which is outside of the scope of this study.

Finally, the choice of a constant mass flux and Stokes number for the pebbles
uniquely defines the pebble surface density, 
\begin{align}
  \Sigma_{\rm peb} = \frac{F_{\rm peb}}{2\pi r v_r} 
  \propto \frac{F_{\rm peb}}{\tau_{\rm f}} \left( \frac{H}{r} \right)^{-2} r^{-1/2} \, ,
\end{align}
where $v_r \approx -2 \tau_{\rm f} \eta  v_{\rm K} $ is the radial velocity of the pebbles, assuming the Epstein drag regime with $\tau_{\rm f} \lesssim 1$. 
Here, $\eta$ is a pressure-gradient parameter defined as $\eta = -0.5 (H/r)^2 (d\ln P/d\ln r)$.
The choice of a constant stokes number thus also has the desirable property
that the dust(pebble)-to-gas ratio ($ \Sigma_{\rm peb}/ \Sigma_{\rm g}$) is
constant with orbital radius and constant in time. 
Therefore, there is no forced pile-up of pebbles anywhere in the disc.

\paragraph{Pebble accretion}
For each body we determine the pebble accretion rate, with a prescription that
is described in detail in Appendix\,\ref{app:pa1}, \ref{app:pa2} and \ref{app:padrift}.
Therefore, the growth of the embryos we model is well-covered by a prescription
that spans the Bondi (drift-dominated) to Hill (shear-dominated) accretion
regimes \citep{Lambrechts_2012}. 
Importantly, we take the dependency of the  eccentricity and inclination on the pebble accretion rate into account (Appendix\,\ref{app:pa1}). 

\paragraph{Filtering and pebble isolation mass}
We reduce the radial flux of pebbles inwards of a body by the fraction the body
accreted. Because of this ``pebble filtering'', the inner bodies see a reduced pebble flux  \citep{Lambrechts_2014b, Guillot_2014, Morby_2015}. 

When sufficiently massive, planets can get isolated from pebbles \citep{Morby_2012, Lambrechts_2014a}. 
The gravitational perturbation of the gas by the embryo creates a pressure bump outside of the orbit of the body trapping the inwards drifting pebbles. 
This isolation mass can be expressed as
\begin{align}
  M_{\rm iso} \approx 10 \times \left( \frac{H/r}{0.04}\right)^3 {\rm M}_{\rm E} \, ,
\end{align}
which is a reasonable approximation in low-viscosity discs \citep{Lambrechts_2014a}. 
More detailed prescriptions that more-or-less agree have recently become
available \citep{Bitsch_2018, Picogna_2018, Ataiee_2018}.
When a body reaches this mass we halt pebble accretion $\dot M_{\rm peb} = 0$
and stop the flux of pebbles to bodies in interior orbits.

\section{Simulation results: before gas dissipation}

\begin{figure}[t!]
  \centering
  \includegraphics[width=8.8cm]{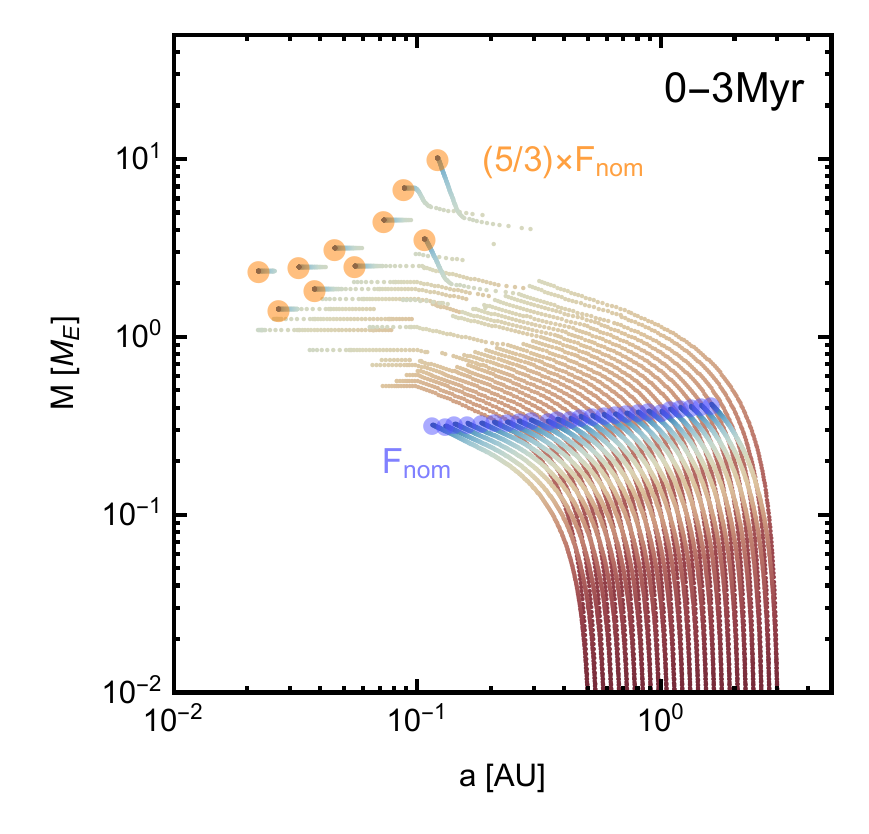} \\
  \caption{
  Embryo mass and semi-major axis for each embryo as function of time, for two
different global pebble mass fluxes. 
  Growth tracks resulting in the final system indicated with blue circles correspond to \texttt{runf3-1}. 
  The system indicated by orange circles corresponds to \texttt{runf5-1}, which
experienced a factor $1.7$ times higher pebble flux.
  }
  \label{fig:ma_run1}
\end{figure}

\subsection{Terrestrial systems}
When the pebble flux is low, such that the total mass in pebbles entering the
inner disc is less than $\approx 110$\,M$_{\rm E}$, the initial embryo
population does not grow beyond approximately $5$ Mars mass (blue circles in
Fig.\,\ref{fig:ma_run1}). 
Mass growth is driven by pebble accretion. Collisions between embryos are rare, because the type-I torques on the planet damp eccentricities and inclinations.
The embryos only experience a moderate amount of migration. 
After $3$\,Myr of evolution the inner embryo resides around $0.1$\,AU, slightly
inwards of the current orbit of Mercury. 
Because all embryos increase in mass at relatively
similar rates, with only slightly higher accretion rates for the outer embryos,
there is no substantial convergent migration.
As a result, embryos grow orderly by smooth pebble accretion up to a
few Mars mass, with little migration. This will change when we consider higher
pebble mass fluxes.

\begin{figure}[t!]
  \centering
  \includegraphics[width=8.8cm]{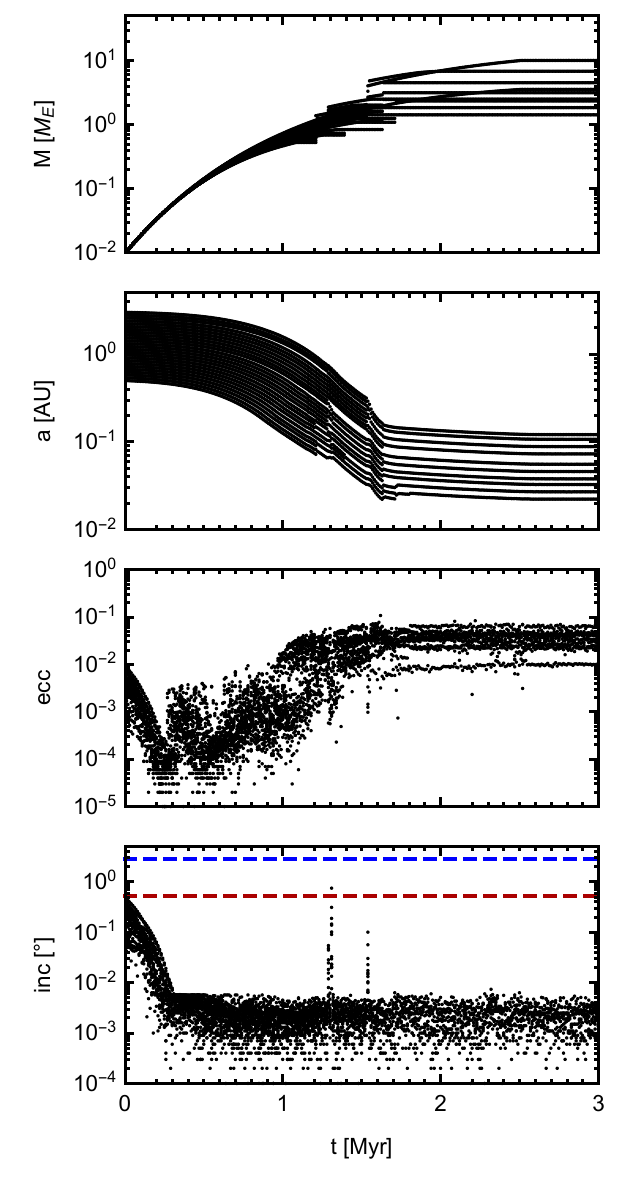} \\
  \caption{
  Example of the time evolution in the gas phase of the disc (\texttt
{runf5-1}).
  Top panel shows the evolution of the mass of the embryos. 
  The next panel shows the decay of their orbits through type-1 migration, which only comes to a halt as embryos approach the inner edge.
  The bottom two panels show the evolution of the eccentricity and inclination.
  In the last panel, the blue line corresponds to the opening angle of the gas scale height with respect to the midplane, and similarly the red curve shows the opening angle of the particle layer.
  }
  \label{fig:tevolgas}
\end{figure}

\subsection{Super-Earth systems}
A larger pebble mass flux that deposits more than $190$\,M$_{\rm E}$ in the inner disc drastically changes the final masses and orbits of the embryos.
When the disc dissipates after $3$\,Myr, embryos are located in short period
orbits within approximately 0.1\,AU, some having been pushed inward of the trap
by larger exterior embryos which filtered the pebble flux.
The embryos now reach super-Earth-like sizes (Fig.\,\ref{fig:ma_run1}).
We illustrate the evolution of the embryos in the gas disc
phase in more detail in Fig.\,\ref{fig:tevolgas}, for a nominal simulation (\texttt{runf5-1}).
In the first Myr of disc evolution, embryo growth proceeds orderly by sweeping
up pebbles.
The embryos develop a small mass spread, which is caused by pebble
accretion rates being higher on the slightly more massive embryos. 
However, this picture of smooth growth changes when embryos reach Earth-like
sizes.
Then, embryos rapidly migrate to the disc edge where their inwards drift comes to a halt.
There the piled-up embryos get dynamically excited. 
The lower-mass embryos generally see the largest eccentricity increase, which results in reduced pebble accretion rates and the suppression of their growth (Fig.\,\ref{fig:mf_run1}).
Larger embryos can accrete more efficiently and see their growth boosted by mutual merging events. 
When embryos grow larger than approximately $5$\,M$_{\rm E}$, pebble filtering
becomes important: the outer embryos accrete at high rates, which reduces the
flux of pebbles to the inner embryos.
Consequently, the outer embryos can migrate inward faster than the less massive inner embryos and overtake them. 
Then, towards final times, the outer embryos accrete the
remaining pebble flux while remaining close to the trapping radius, as can be observed by the final steep growth curves in Figure\,\ref{fig:ma_run1}.
Even larger embryos reach the pebble isolation mass of $10$\,M$_{\rm E}$, 
which then shuts down the pebble flux to inner embryos.
In the last Myr before disc dissipation, the embryo system typically evolves little. 
Besides the dynamical excitation of the embryos, this is mainly the result of
the diminished pebble flux, because of mutual pebble filtering and the general
time decay of the global pebble flux. 

\begin{figure}[t!]
  \centering
  \includegraphics[width=8.8cm]{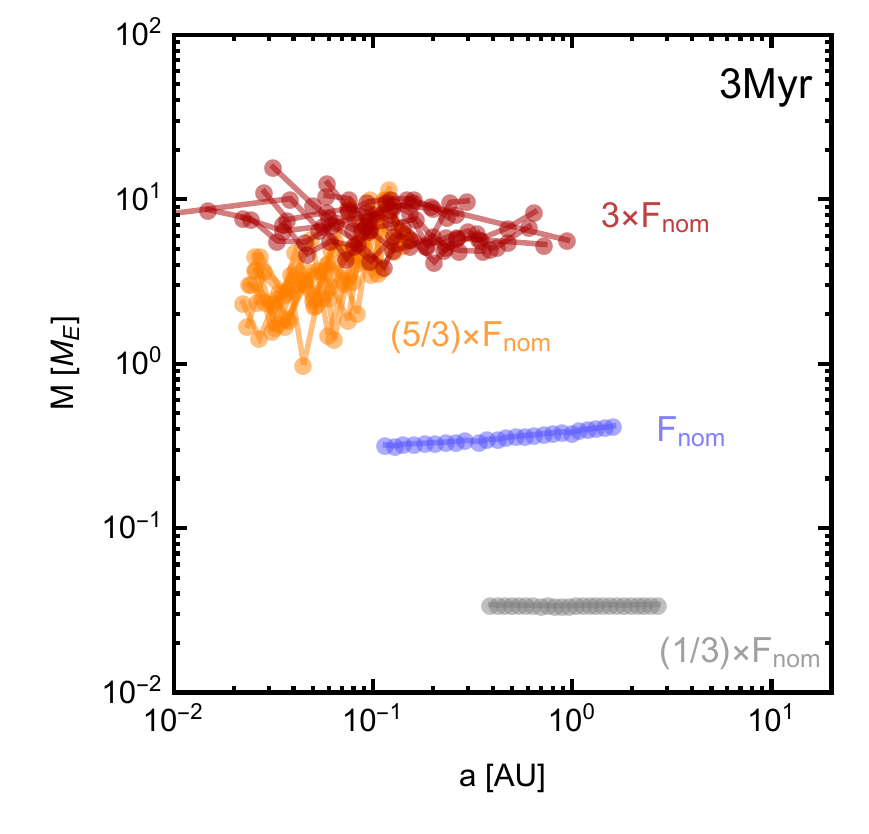}
  \caption{
Final system architecture, given as the embryo mass versus orbital distance, at disc dissipation
(3\,Myr). 
Different colors represent different pebble mass-fluxes, blue, orange, red,
experienced a respectively 3,5,9 times higher mass flux than the gray systems.
  }
  \label{fig:3Myrsys}
\end{figure}

\begin{figure}[t!]
  \centering
  \includegraphics[width=8.8cm]{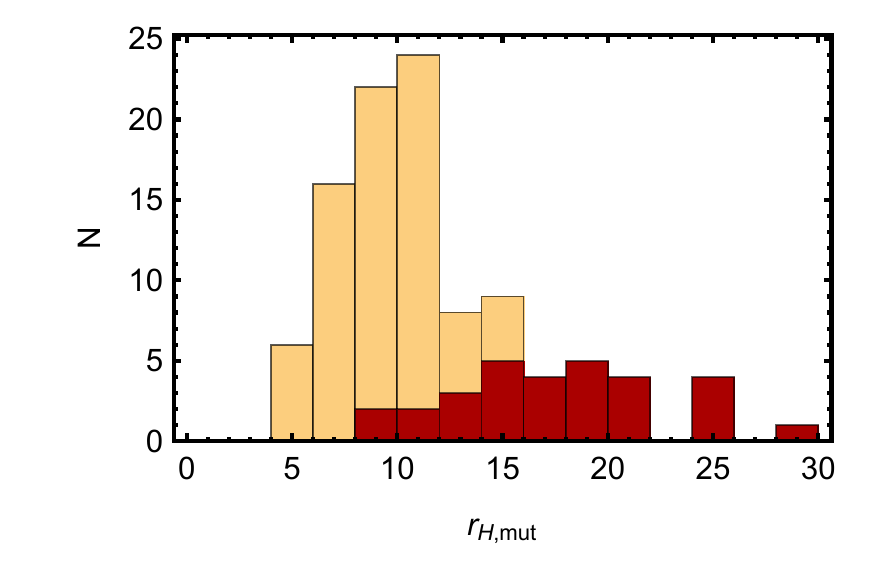} \\
  \caption{
   Planet separation, expressed in mutual Hill radii, between neighbouring planet pairs pairs for runs \texttt{runf5}, after 3-Myr year of evolution in the gas disc (yellow) and after 100\,Myr of subsequent evolution (red). The most closely spaced planet pairs do not survive.
  }
  \label{fig:mHr}
\end{figure}

\begin{figure}[t!]
  \centering
  \includegraphics[width=8.8cm]{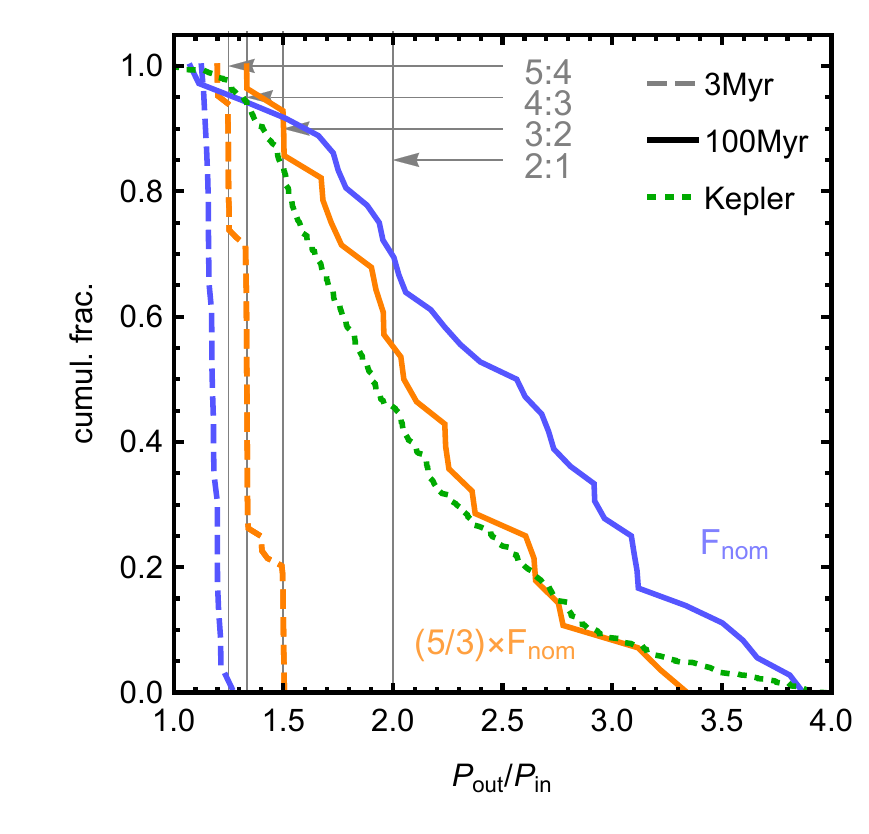} \\
  \caption{
  Cumulative distribution of neighbouring planet pairs. 
  Dashed curves show the situation after the end of the gas disc phase
($t=3$\,Myr), for both simulations suites \texttt{runf3} (blue) and
\texttt{runf5} (orange). 
  The full lines show the situation after $100$\,Myr of
additional evolution.
  For the super-Earth systems most resonant pairs do not survive the post-gas
phase.
  The vertical gray thin lines show the period ratios corresponding to, from
left to right, the $5{:}4, 4{:}3, 3{:}2, 2{:}1$ first order mean motion
resonances.
  For comparison, the green short-dashed curve shows the observed period ratio
distribution of the Kepler multi-planet systems, limited to a maximal period
ratio of $P_{\rm out}/P_{\rm in}=4$. 
  }
  \label{fig:PP}
\end{figure}

For the formation of super-Earths the gravitational interactions between
embryos are important. 
Therefore we ran $10$ simulations (suite \texttt{runf5}) to capture the nominal outcome.
The final systems, after 3\,Myr of evolution in the gas disc, are shown in
orange in Figure \ref{fig:3Myrsys}.
Embryos are larger than 1 Earth mass in size,  and typically do not grow larger than about $10$\,M$_{\rm E}$.
We find that the super-Earth cores are located between the inner edge and
approximately $0.1$\,AU. 
The embryos are found to be in relatively compact configurations, which is
characteristic for systems that evolved through orbital migration and
experience tidal damping of the eccentricity and inclination \citep{Ogihara_2015}.
Most neighbouring planet pairs are separated within $13$ mutual Hill radii\footnote{
The mutual Hill radius is defined as 
$r_{\rm H, mut} = \frac{a_1 + a_2}{2} \left( \frac{m_1+m_2}{3 M_\odot} \right)^{1/3}$. 
Here, $m_1,m_2$ are the masses and $a_1,a_2$ the semi-major axes of the planet
pair.
} from each other, as can be seen in Fig.\,\ref{fig:mHr}.
Also, as expected, orbital migration caused many planets pairs to fall close
to first-order mean-motion resonances (Fig.\,\ref{fig:PP}).
A significant fraction turn out to appear to be near the $4:3$ mean-motion
resonance. 

For completeness, we also explored simulations with higher pebble flux
(suite \texttt{runf9}).
Predictably, we find the embryos grow to larger sizes, in the $5$ to $10$\,M$_{\rm E}$ regime (red circles in Fig\,\ref{fig:3Myrsys}).
Faster growth leads to an earlier migration of the embryos towards the inner edge.
Generally, we observe a more violent dynamical evolution, where embryos can
even be excited to orbits with inclinations above the pebble midplane, which
strongly reduces pebble accretion rates.
The final systems show a wider orbital spread, out to approximately $1$\,AU.
The majority of planet pairs are now close to first-order $j/(j+1)$ mean motion
resonances with low j ($j=1,j=2$). 
This is because more massive embryos of are more likely to get trapped in more
distant first-order resonances \citep{Ogihara_2013}. 
This causes the systems to be more extended compared to the systems with lower
pebble flux of \texttt{runf5}.

\subsection{Understanding the dependency on the pebble flux}
The steeper than linear relation between the pebble flux and final embryo mass
is driven by pebble accretion, and steepened further by growth through embryo
mergers.
Indeed, in our disc model embryos typically accrete in the 3D regime, where the accretion radius is smaller than the pebble scale height
$r_{\rm acc} \lesssim H_{\rm p}$. 
In the so-called strong-coupling limit of the Bondi and Hill branch \citep[][ see also Apppendix \ref{app:pa1}]{Lambrechts_2012}, one then finds embryo growth rates of the form
\begin{align}
  \dot M_{\rm p} \approx
  \frac{1}{4\sqrt{2\pi}} \frac{1}{\eta} 
  \left(\frac{H_p}{H} \right)^{-1}
  \left(\frac{H}{r} \right)^{-1}
  F_{\rm peb} \frac{M_{\rm p}}{M_\odot} \,,
  \label{eq:3Dacc}
\end{align}
which implies that the embryo mass $M_{\rm p}$ has an exponential dependency on the mass flux $F_{\rm peb}$ 
\citep[][see Apppendix \ref{ap:3dstrong} for a derivation]{Ormel_2010,Lambrechts_2012,Ida_2016,Ormel_2017b,Ormel_2018,Lin_2018}
Therefore, a change of a factor $2$ in pebble mass flux can lead to almost an
order of magnitude change in final embryo mass. 
This steep dependency breaks down when embryos masses become large, get more excited and mutual filtering and isolation become important.

We summarize the final systems at the time of disc dissipation, for different pebble fluxes, in Fig.\,\ref{fig:3Myrsys}, which also includes at control simulation with a low pebble flux (\texttt{runf1-1}).
Combining the results obtained in the gas disc phase also allows us to express the mean embryo mass, as function of the cumulative pebble flux through the disc in Figure \ref{fig:fluxvsM}.

\section{Simulation results: after gas dissipation}

\subsection{Terrestrial systems}
We now follow up on the evolution of the terrestrial embryos, after the gas
disc has dissipated, for an additional $100$\,Myr. 
First, we consider the evolution of the set of our terrestrial embryos that
grew to approximately a few Mars masses, while in the gas disc (suite
\texttt{runf3}). 
After the disc dissipates, all the small embryo chains become unstable. 
Bodies grow substantially over the next tens of Myr through mutual mergers,
just as expected from classical terrestrial planet simulations as reviewed in
\citet{Morbidelli_2012b}.
Their formation history is thus similar to the formation of the Earth through
giant impacts \citep{Chambers_2001,Obrien_2006,Raymond_2009,Izidoro_2014,Jacobson_2014b}.

\begin{figure}[t!]
  \centering
  \includegraphics[width=8.8cm]{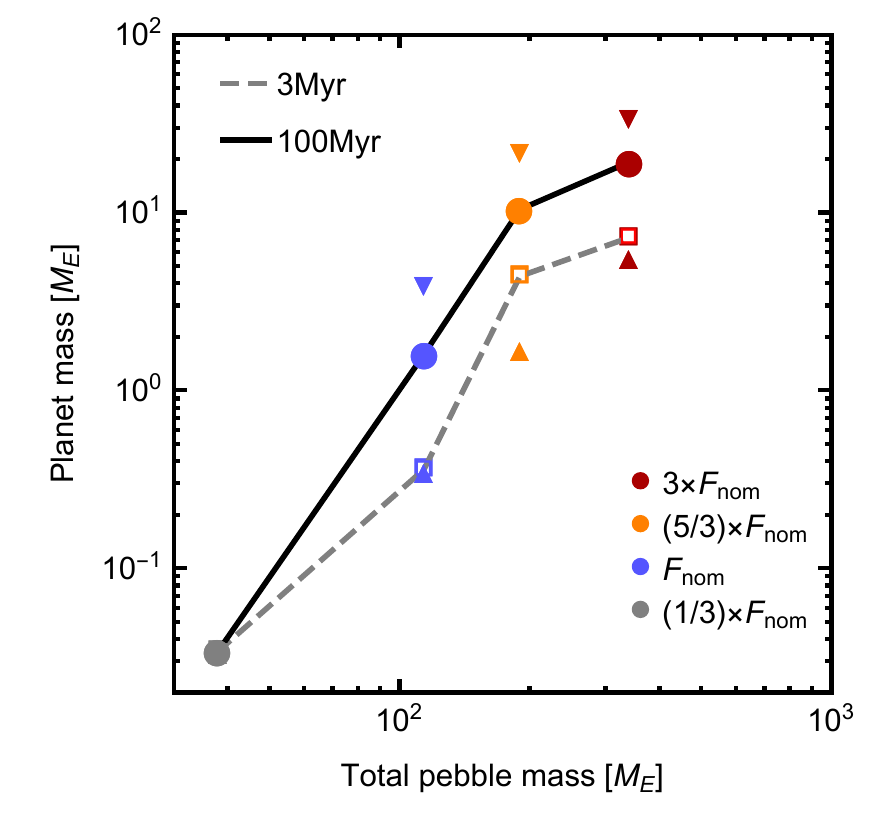}
  \caption{
Final planet mass as function of the integrated pebble flux.  
Circles, connected with the full black curve, give the mean planet mass across
a suite of runs with same pebble flux, while triangles represent the minimal
and maximal embryo mass across all runs with the same pebble flux.
The squares, connected by the gray dashed line, give the mean embryo mass at
the time of disc dissipation.
  }
  \label{fig:fluxvsM}
\end{figure}

\begin{figure}[t!]
  \centering
  \includegraphics[width=8.8cm]{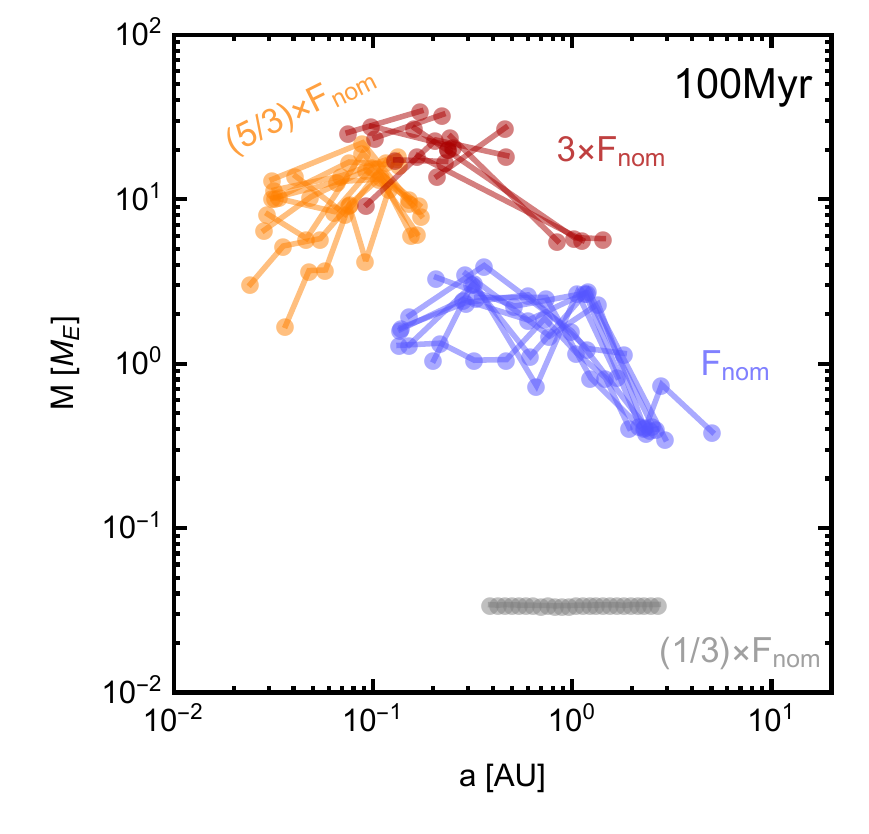}
  \caption{
Final system architecture, given as planet mass versus orbital distance, after $100$\,Myr of evolution. 
Color coding is similar as in Fig.\ref{fig:3Myrsys}.
Scattering, collisions and ejections generally reduce the number of planets per system and set the final orbital architecture of the system.
  }
  \label{fig:100Myrsys}
\end{figure}

The final systems are shown in blue in Fig.\,\ref{fig:100Myrsys} (suite \texttt{runf3-C}). 
Planets grow up to approximately $4$\,M$_{\rm E}$. 
This can thus be considered to be the upper mass limit for what we would
classify as a terrestrial planet. 
Most of the planets are spread between $0.1$\,AU and a few AU, corresponding to the outer edge of the original embryo disc. Some get placed in wider orbits of up to $10$\,AU.
On average about 5 planets remain in the system. 
The mean eccentricities are relatively high, around $e=0.1$, but we do not include dynamical friction by planetesimals. 

These simulated planetary systems share some similarities with the terrestrial planets in the Solar System.
Generally, the most massive planets are found between $0.5$ and $1$\,AU and less massive planets are present in the interior and exterior of that zone.
These are the outcome of unfortunate embryos that get kicked out of this central region, after which their growth comes to a halt. 
This is a generic outcome of terrestrial planet simulations where embryos are
initially confined in a narrow annulus. 
Such an embryo configuration has previously been proposed to explain the origin
of Mars as a stranded embryo \citep{Hansen_2009,Raymond_2009,Walsh_2011,Morbidelli_2012b,Jacobson_2014b,Walsh_2016}.  

The aim here is not to recreate the exact configuration of the terrestrial
planets in the Solar System. 
However, we did experiment and found that a moderately reduced integrated mass
flux, of about $60$\,M$_{\rm E}$, leads to planets with masses in the Venus to
Earth-mass regime.
Finally, in our control simulation with the lowest pebble flux that we
considered (\texttt{runf1-1}), the embryos are sufficiently small at disc
dissipation so that they can avoid mergers at later times \citep{Iwasaki_2006}. Therefore, these
sub-Mars embryos remain small (gray points in Fig.\,\ref{fig:100Myrsys}).

\subsection{Super-Earth systems}

\begin{figure*}[t!]
  \centering
  \includegraphics[width=8.8cm]{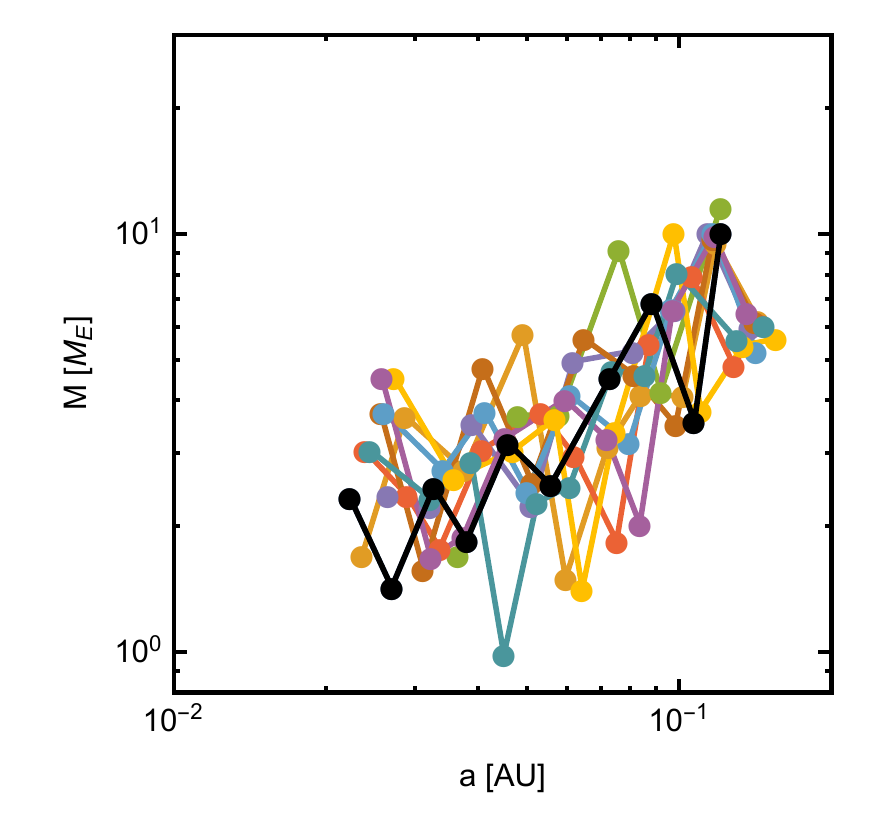}
  \includegraphics[width=8.8cm]{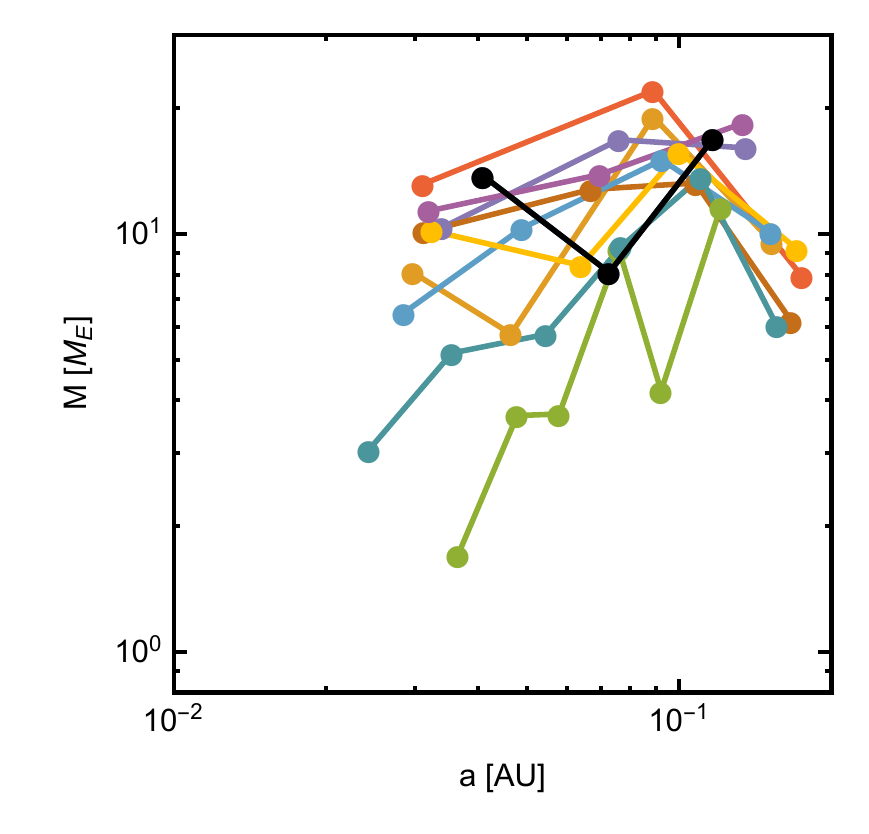}
  \caption{
  The left panel shows 10 systems after $3$\,Myr of evoluton in the
protoplanetary disc (\texttt{runf5}).
  The right panel shows the same systems, after $100$\,Myr of evolution
(\texttt{runf5-C}). 
  Only one remained stable, here shown as the light green system, while the
other systems all underwent a post-gas instability. 
  These latter systems contain fewer, but more massive planets and are less
strongly size-sorted with orbital distance.
  }
  \label{fig:finalsystems}
\end{figure*}

\begin{figure}[t!]
  \centering
  \includegraphics[width=8.8cm]{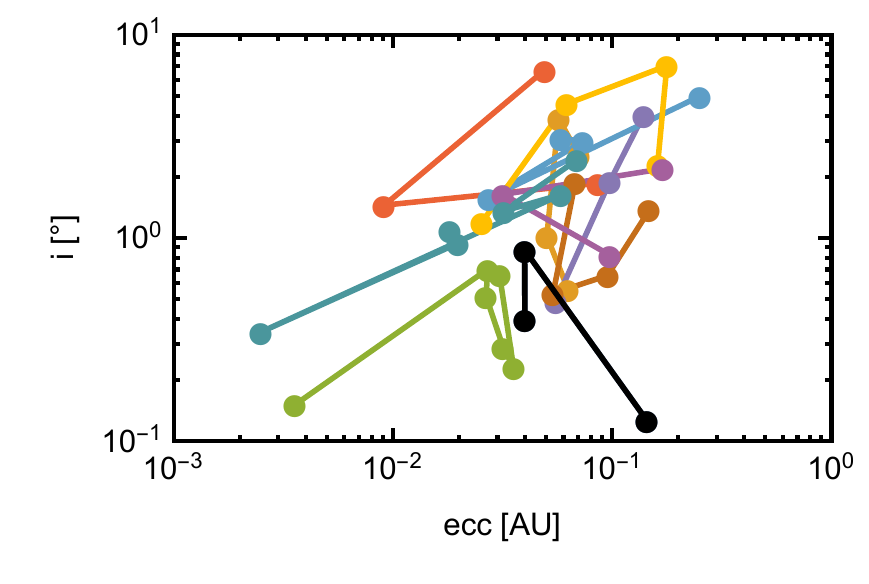}
  \caption{
  Eccentricity and inclination for the final systems, after $100$\,Myr of
evolution (\texttt{runf5-C}).
  Lowest values for eccentricity and inclination are found for the stable resonant system. Generally, lower number systems have higher eccentricities and inclinations.
  }
  \label{fig:eifinalsystems}
\end{figure}

The larger super-Earth-like embryos, with shorter periods (suite
\texttt{runf5}), undergo a different evolution after the disc dissipates,
compared to the terrestrial embryos.
Typically, we find that the removal of the gas disc renders the embryo chain
unstable. 
Then, within usually the first few $10$\,Myr of evolution, embryos merge, or
collide with the central star, and settle in their final configuration.
These systems, after a $100$\,Myr of additional evolution after disc
dissipation (suite \texttt{runf5-C}), can be inspected in
Fig.\,\ref{fig:100Myrsys} (orange points).
The final planetary masses have increased and range from about 1\,M$_{\rm E}$ to about 20\,M$_{\rm E}$. 
In those systems that undergo a post-gas instability, growth is more efficient for the inner embryos.
This can be seen in Fig.\,\ref{fig:finalsystems}, where the mass and orbital
location of each system at disc dissipation can be compared with respect to the
final system.
Thus instabilities after disc dissipation can erase the trend of larger mass
objects in wider orbits that builds up through pebble accretion in the gas disc
phase \citep{Izidoro_2017,Ogihara_2018a}.
On average there are about $4$ planets left in the system. 
The final inclinations are within $10$ degree and most eccentricities fall between $e=10^{-2}$ and $e=0.2$ (Fig.\,\ref{fig:eifinalsystems}).

Not all systems become unstable, as the green system in
Fig.\,\ref{fig:finalsystems} shows (\texttt{runf5-3}).
Compared to the unstable systems it keeps its $6$ planets, which have period
ratios between neighbouring planet pairs that are close to mean motion
resonance ($3{:}2, 4{:}3, 3{:}2, 4{:}3, 3{:}2$). 
It also maintains low eccentricities ($e<0.04$) and inclinations ($i<1$
degree).
This one system is however the exception. 
We estimate that more than $90$\% of all systems experience a post-gas
instability, based on a suite of $19$ simulations where only one system remained
stable\footnote{
Unstable systems were re-simulated with a longer gas removal time scale of
$t_{\rm disc}=5$\,Myr. 
In this way we generate $9$ additional systems for the post-gas integrations,
which revealed all these systems to become unstable. 
Thus, this experiment also indicates our results are not very sensitive to the
choice or $t_{\rm disc}$.
}.

The tendency for these systems to go unstable is not surprising. 
We find that when embryos grow by pebble accretion many of the super-Earth pairs
become too closely spaced when they leave the gas disc phase. 
Nearly all pairs with mutual Hill spacing within $13$\,r$_{\rm H,mut}$ merge after the removal of gas \citep{Matsumoto_2012}. 
This can be seen in Fig.\,\ref{fig:mHr}, by comparing the yellow versus red histograms.
Unsurprisingly, post-gas instabilities also destroy most of the resonant period pairs (orange dashed curve in Fig.\,\ref{fig:PP}). 
The high occurrence of post-gas instabilities is thus different from the
earlier results by \citet{Izidoro_2017}.
They found that only half of the super-Earth systems become unstable after disc
dissipation, because in their model, which treats migration but not pebble
accretion, embryos are generally more widely spaced. 
Mass growth by pebbles does not directly change the physical separation between
embryos, but it does change the mutual Hill spacing.
In the new work by \citet{TRILOGY_ANDRE}, which treats pebble accretion, a
similar high probability is found for the break up of resonant chains of
planets with similar masses at disc dissipation.

The super-Earth systems we find seem to be in qualitative agreement with the
observed population of super-Earths detected by the Kepler survey.
Previously, \citet{Pu_2015} noted that the Kepler planets typically have mutual
Hill separations around $12$\,$r_{\rm H,mut}$, close to the minimal separation
required to survive over Gyr-timescales. 
Therefore, they argued that most super-Earth systems of $\lesssim 4$ planets
formed in an initially more planet-dense configuration. 
The formation scenario presented here thus supports this picture.
Moreover, we find that super-Earth pairs are typically not in mean motion
resonance, in agreement with Kepler observations \citep{Lissauer_2011b}. 
The period ratio distribution we find for the simulated super-Earth systems is qualitatively similar to the one for the observed Kepler-planet pairs (green dashed line in Fig.\,\ref{fig:PP}, see also the companion paper by \citealt{TRILOGY_ANDRE}). 
Eccentricities and inclinations appear also to be broadly consistent, with
Kepler systems having \citep[$e< 0.05$, ][]{Xie_2016} 
and low mutual inclinations \citep[$i \lesssim$ $10^{\circ}$,][]{Lissauer_2011b,Johansen_2012,Zhu_2018}.
Furthermore, the inherent multiplicity of a typical Kepler super-Earth systems
has been inferred to be about $4$ \citep[although this is dependent on the
inclination model used, ][]{Johansen_2012,Izidoro_2017,Zhu_2018}. 
This would agree with our simulations which have, on average, $4$ surviving
planets.
Also, recently, \citet{Wu_2018} argued, based on modelling the planetary radius distribution, that the Kepler systems are composed of planets with a characteristic rocky mass of about $8$\,M$_{\rm E}$, with weak orbital radius dependency.
This appears to be roughly consistent with our simulations showing planetary masses do not show a strong dependency on orbital radius and that planets have a mean mass around $10$\,M$_{\rm E}$.
%
For a more quantitative discussion on observational implications, the reader
can consult \citet{TRILOGY_ANDRE}.

We now briefly discuss the evolution of the embryos that formed from the
largest pebble mass flux we considered (suite \texttt{runf9-C}).
Their post-gas growth evolution is largely equivalent to the other super-Earth simulations, where disc dissipation generally triggers a post-gas instability. 
The final masses and orbits of the planets in these systems are shown in Fig.\,\ref{fig:100Myrsys} (red circles).
Given that these runs use a large cumulative pebble mass flux of $350$\,M$_{\rm
E}$ and a disc aspect ratio resulting in a relatively large pebble isolation
mass of $10$\,M$_{\rm E}$, these simulations probe the most massive super-Earth
systems we can conceivably form.
From these results it thus seems implausible that the rocky cores of
super-Earths grow beyond $30$\,M$_{\rm E}$ in mass.

To summarise, the strong dependency of the embryo mass on the pebble flux at
disc dissipation remains reflected in the final planetary masses. 
The relation between the cumulative pebble mass flux and the mean planetary
mass is illustrated in Fig.\,\ref{fig:fluxvsM} (black curve).
A small increase in the pebble mass flux by a factor of $2$ leads to the
formation of super-Earths larger than $10$\,M$_{\rm E}$ in mass, instead of
terrestrial planets in the Earth-mass regime.

\section{Identifying super-Earths and true terrestrial planets}

We have argued that systems of terrestrial planets and systems of
super-Earth systems are distinct in the way they form.
However, since their growth histories cannot be observed, we here summarize,
and attempt to quantify where possible, the observable differences between
these two different types of planetary systems.
We believe these two classes of systems should be distinct, because our
simulations do not argue for planetary systems which could consist of planets
with both terrestrial and super-Earth growth histories.
A caveat here is that we consider here a growth channel based on a single
population of initially similar-sized rocky embryos. Super-Earths may form
outside the ice line and migrate inwards, which could leave behind mixed systems
\citep{Cossou_2014,Raymond_2018,TRILOGY_ANDRE}. 

Systems of super-Earths contain planets more massive than the terrestrial
planets. 
However, a simple mass threshold is not sufficient to distinguish between a
terrestrial or super-Earth growth history.
A system with planets in the approximately $1$ to $5$\,M$_{\rm E}$ regime can
have formed in either growth mode (Fig.\ref{fig:fluxvsM}).
Only when planets more massive than approximately $5$\,M$_{\rm E}$ are present,
the system is likely composed of true super-Earths.
We also note that we do not find a strong relation of the planetary mass with orbital radius for the super-Earth systems, while there is a tendency for the outer terrestrial planets to decrease in mass with orbital radius (Fig.\ref{fig:100Myrsys}).
These findings are also sensitive to the pebble isolation mass, which in our
disc model is about $10$\,M$_{\rm E}$. Discs with smaller isolation masses
would further limit growth by pebble accretion \citep{Bitsch_2018}.

We can also consider the difference in the orbital architecture \citep{Raymond_2008}.
Our super-Earth systems are typically more concentrated towards the disc edge, around $0.1$\,AU.
The terrestrial systems on the other hand trace their origin location better
and stretch out beyond distances of $1$\,AU.  
Also, systems found to be in a resonant chain point to a super-Earth-like
formation history.

When the composition of a planet can be determined, it can also be used to help
distinguish terrestrial from super-Earth systems.
Because our terrestrial planets form from small Mars-sized embryos, the final planets cannot have significant gaseous envelopes. 
The envelope mass fraction that could be expected from outgassing alone is in the range of at most a few percent \citep{Rogers_2011}. 
Conversely, the super-Earth planets can have significant gaseous envelopes, but
not necessarily so.
In this work we have not explored gas accretion onto embryos. 
Isolated planets exceeding approximately $1$\,M$_{\rm E}$ can accrete about 10\,\% of their total mass in gas during the disc phase \citep{Lee_2014,Ormel_2015,LL_2017}.
However, our simulations also reveal that planets can experience mutual collisions, both in and after the gas phase. 
Therefore, it may be possible that part of the original envelopes are lost
\citep{Liu_2015,Schlichting_2015}. 
Moreover, close-in planets may loose their envelopes through stellar irradiation \citep{Baraffe_2006, Owen_2017,Carrera_2018}.
Finally, some of the largest cores could even undergo runaway gas accretion and in this way escape the super-Earth class, by becoming gas giants \citep{TRILOGY_BERT}.
Clearly, the role of gas accretion is an area for further study. 
Nevertheless, we can conclude that when a significant gas envelope is present
around a planet, the accretional history must have been super-Earth-like.

Taken together, we argue here that one ideally should consider the masses,
orbital architecture and composition of the planets in a system as a whole, in
order to observationally distinguish terrestrial from super-Earth systems.
In this way we can conclude that a good terrestrial planet candidate is smaller
than $5$\,M$_{\rm E}$ and that it is part of an extended non-resonant multiple
planet system of similarly small planets. 
Moreover, these planets should have no gaseous envelopes, or small envelopes that do not exceed a few percent of the total planetary mass.
Therefore, it is at this point in time not yet clear if a true terrestrial planet has been observed outside of the Solar System.

\section{Discussion}
\subsection{Pebble mass reservoir}

In this work, we considered the integrated mass flux of pebbles through the
inner disc as a free parameter.
The distribution of the total mass available in pebbles in protoplanetary
discs around solar-like stars is observationally poorly constrained.
ALMA surveys of Myr-old star-forming regions argue for dust masses between
crudely $\sim$$1$ and $\sim$$100$\,M$_{\rm E}$, as inferred from (sub-)mm
emission around Class II sources \citep{Ansdell_2017,Dullemond_2018}.
However, such measurements assume that the emission is optically thin, which
may not be the case at mm wavelengths. 
For example, longer wavelength measurements of the young HL Tau system argue
for a total dust reservoir of $300$ to $1000$\,M$_{\rm E}$
\citep{Carrasco_2016}, a factor $3$ times larger than inferred from ALMA
measurements \citep{Pinte_2016}.
Additionally, these observations miss all mass located in larger sizes.
Indeed, given the uncertain ages of stars in starforming regions, it is
plausible that a large fraction of the mass reservoir is already locked
up in growing planets \citep{Najita_2014,Manara_2018,Johansen_2018}.
In our simulations, for example, the embryos grow rapidly in a short time interval of approximately $1.5$\,Myr. 
This may then also explain why the dust masses inferred around even younger
Class 0 sources are substantially higher, with median dust masses around
$250$\,M$_{\rm E}$, compared to the above mentioned Class II sources
\citep{Tychoniec_2018}.


Because super-Earths appear around approximately a third of solar-like stars
\citep{Zhu_2018}, without being strongly
dependent on stellar metallicity \citep{Buchhave_2012}, it appears the pebble
flux for super-Earths systems is commonly available.
The exact value of the required pebble mass (we find here approximately
$190$\,M$_{\rm E}$) depends on the filtering efficiency $\dot M_{\rm
p}/F_{\rm peb}$ (Eq.\,\ref{eq:3Dacc}). The latter needs to be numerically determined (Appendix \ref{app:pa1}).
Recent numerical efforts argue for approximately a factor $4$ higher filtering
efficiencies, which would reduce the required mass in pebbles, but efficiencies
decrease again with increasing turbulence \citep{Xu_2017,Ormel_2018}.
We also note that filtering factors are higher around low-mass stars
\citep[Eq.\,\ref{eq:3Dacc}, ][]{Ormel_2017} possibly explaining why super-Earth
occurrence rates remain high around such small stars \citep{Mulders_2015}.

To form the terrestrial planets, a pebble mass reservoir of about $110$\,M$_{\rm
E}$, within less than a factor $2$, is required.
This relatively narrow range in the pebble flux appears to indicate that the
formation of terrestrial planets may be less common than the formation of
super-Earth systems.
However, as we argued above, we do not know the true distribution of the
total pebble masses in protoplanetary discs, which may often fall in the 
range that produces terrestrial planets.
An additional complication is that the mass flux past the ice line would be
modified when outer giant planets are present that can filter and even halt the
flux of pebbles \citep{Lambrechts_2014a}.
For example, in the context of the Solar System, when the core of Jupiter
reached pebble isolation in the outer disc it should have halted the mass flux
of pebbles to the inner disc, which could then have limited the growth of inner
planetary embryos to Mars-mass \citep{Morby_2015}.

In general, the early formation of pebble-filtering giant planets in
wide orbits could suppress the formation of close-in super-Earths.
However, \citet{Zhu_2018b} argue, based on statistical grounds, that there
appears to be a correlation between close-in super-Earths and wide-orbit giant
planets, although radial velocity surveys show this line of evidence may not
yet be conclusive \citep{Barbato_2018}. 
If this correlation indeed holds, it implies that the Solar System
configuration with terrestrial planets and wide orbit gas giants is rare.
It could then mean that typically gas giants form late, close to disc
dissipation, which would minimize their effect on the pebble flux.
Or, alternatively, hint that these super-Earths do not follow the growth path
we investigated in this study where rocky cores are grown inside the ice line.
Indeed, the companion paper by \citet{TRILOGY_BERT} shows that the formation of
giant planets outside the ice line does not appear to generally prevent smaller
icy super-Earth-sized planets from migrating inwards.

In summary, we believe that we have invoked plausible pebble mass reservoirs.
Future work is needed to more precisely quantify the required mass needed to
form observed planetary systems and the role of planets in the outer disc in
shaping the pebble flux.
Irregardless, the strong dependency that we have highlighted between the pebble
flux and the type of planetary system thats is formed should be robust.

\subsection{Summary of simplifying assumptions}

In order to model the growth of the planetary embryos, we took into account
pebble accretion, planetary migration and gravitational interactions with the
help of an N-body code.
We found that, together these three processes shape the final planetary systems.
However, in this study we made several simplifying assumptions that deserve to be studied in more detail.

We considered a simple model for the gas disc and its inner edge, which 
only allows for inwards type-I migration of embryos. 
In our particular disc model, equal-sized embryos migrate with the same migration timescale independent of orbital radius, which does not necessarily hold in disc models with steeper density gradients. 
However, for the low pebble flux cases migration is little relevant, while for
the high pebble flux cases migration is convergent as the outer embryos tend to
grow larger than the inner ones and because there is a planet trap at the inner
edge of the disc.

We used a simplified prescription of the disc edge, which also ignored the role of the exterior silicate sublimation line. 
However, the inner edge is critical to prevent the super-Earth embryos from migrating into the star. Therefore, our work would benefit from an improved physical description of the disc edge \citep{Flock_2017}, while also treating the particle size evolution through coagulation and sublimation \citep{Ros_2013,Schoonenberg_2017}.

We also only considered a single population of close-in embryos seeds located inside of the ice line. 
The initial embryo masses matter because, for a single embryo grown by pebble
accretion, the final mass after exponential growth is linearly dependent on the
initial embryo mass (Eq.\,\ref{eq:3Dacc}).
This remains approximately true for multiple embryos when they are small enough to ignore pebble filtering. However, this breaks down around larger embryos or when systems become sufficiently dense to cause dynamical excitation.
Ultimately, where and when the first embryos emerge is uncertain \citep[for example, a different time-dependent embryo emergence is explored in][]{Ormel_2017}.
We did not further explore the initial embryo distribution here.

While we assumed the embryos seeds to exist only within the iceline, the
companion papers \citet{TRILOGY_ANDRE} and \citet{TRILOGY_BERT} consider embryo
seeds also located beyond the snowline, as well as the effects of a migrating
snowline, in more detailed disc models with migration traps
\citep{Bitsch_2015a}. 
They find that icy embryos growing beyond the snowline generally
migrate inwards. 
This would disrupt the formation of systems of rock-dominated cores as outlined
in this work.
If super-Earth cores are indeed dominantly rocky in composition
\citep{Owen_2017,Lopez_2017}, this remains an open problem.

\section{Conclusions}

We studied the migration and growth of rocky embryos within
the ice line around a solar-like star. 
Additionally, we followed their subsequent post-gas disc evolution.
In the gas phase, we find the pebble mass flux strongly regulates the final masses of the embryo.
A factor of two difference in the pebble mass flux can result in a change from Mars-sized embryos to larger than Earth-mass ones.
This strong difference is caused by the fact that, if embryos become
Earth-sized, they start rapidly migrating and become highly efficient in
accreting pebbles.
These planets pile up close to the disc edge and their growth by pebble
accretion is limited by the pebble isolation mass.
After the gas disc dissipates, the smaller Mars-sized embryos grow through
mutual mergers to planets in the Earth-mass regime, forming terrestrial
planets, like those in the Solar System. 
The larger super-Earth planets typically experience instabilities in the
post-gas evolution, because the combination of pebble accretion and migration
left behind compact systems with planets in closely-spaced resonances, with
small separations as measured in their mutual Hill radii.
In this way the systems are typically dislodged from their resonant chains.

The largest planets that form in the terrestrial mode can become as massive as
approximately $5$\,M$_{\rm E}$. Although, to discriminate between terrestrial
and super-Earth systems from an observational viewpoint, we argue a mass threshold is not sufficient. Fortunately, the orbital architecture and the presence of gaseous envelopes can be used to separate the two classes of planetary systems.

The pebble mass flux was chosen to be a free parameter in this study. Further
observational constraints on the distribution of initial dust disc masses are
needed in order to asses if this formation model is in agreement with observed
exoplanet occurrence rates. 
This also requires continued work on the precise determination of the pebble accretion efficiency \citep{Liu_2018,Ormel_2018}.
The complete evolution of the gas disc, pebble component and the embryos across
the disc remains a key point for further exploration.  Companion papers
\citet{TRILOGY_ANDRE} and \citet{TRILOGY_BERT} expand this study beyond the ice
line. There, outer planets can reduce and halt the pebble flux. Moreover, icy
cores can migrate across the ice line.  

To conclude, we have shown here two growth modes, regulated by only the radial
pebble mass flux.
When the pebble flux is sufficiently high, 
we no longer form terrestrial systems from rocky embryos, but 
instead resonant chains of super-Earths by migration-assisted growth. 
These compact systems typically become unstable after disc dissipation,
leaving behind non-resonant systems of approximately four super-Earths with
eccentricities and inclinations consistent with observed values.

\begin{acknowledgements}
M.L.\,thanks Masahiro Ogihara, Yasunori Hori and  Eiichiro Kokubo for
stimulating discussions at DTA symposium VIII.
The authors are grateful for the constructive feedback by an anonymous referee.
M.L., A.M.\,and S.R.\,were supported by ANR through project MOJO (Modeling
the Origin of JOvian planets, ANR-13-BS05-0003-01).
S.J.\,and A.M.\,were supported by the European Research Council
(ERC) Advanced Grant ACCRETE (contract number 290568).
A.J.\,is grateful for support from the KAW Foundation (grant 2012.0150), the European Research Council (ERC Consolidator Grant 724687-PLANETESYS) and the Swedish Research Council (grant 2014-5775).
B.B.\,thanks the European Research Council (ERC Starting Grant 757448-PAMDORA)
for their financial support.
A.I.\,gratefully acknowledges financial support from FAPESP via grants \#16/19556-7 and \#16/12686-2.
\end{acknowledgements}

\bibliographystyle{aa}        
\bibliography{references}     

\appendix

\section{Pebble accretion prescription} 
\label{app:num}

\subsection{Accretion radius}
\label{app:pa1}

The pebble accretion radius is calculated as function of the relative velocity between pebbles and the embryo, and the drag force the particle feels.
We verified the prescription described below against our 2-body integrations
that include drag, performed in the shearing sheet \citep{Ormel_2010,
Lambrechts_2012,Johansen_2015,Liu_2018}.
Below, we follow the terminology of \citet{Lambrechts_2012}. 

Massive bodies accrete in the Hill regime, where Keplerian shear largely determines the accretion rate. 
Lower mass bodies typically accrete in the so-called strong Bondi-branch, where particles are well-coupled to the gas, but approach velocities are now dominated by the sub-Keplerian velocity of the gas with respect to the embryo.

In these $2$ regimes, accretion relies on the friction time across the
accretion radius to be shorter than the deflection time. 
This then sets an effective accretion radius.
The friction time is given by 
\begin{align}
  t_{\rm fric} = \frac{\rho_{\rm s}R}{\rho c_{\rm s}} \,,
\end{align}
where $R$ and $\rho_{\rm s}$ are respectively the radius and density of the particle.
The deflection time is expressed as 
\begin{align}
  t_{\rm def} = \frac{v_{\rm acc} r_{\rm acc}^2}{GM}\,.
\end{align}
Here, $M$ is the mass of the embryo and $v_{\rm acc}$ is the accretion velocity.
The latter is dependent on the accretion radius through 
\begin{align}
  v_{\rm acc} = v_{\rm rel} + \frac{3}{2} r_{\rm acc} \Omega_{\rm p}
\end{align}
The relative velocity of the embryo with respect to the pebbles orbiting in a sub-Keplerian gas disc is given by
\begin{align}
  v_{\rm rel}^2 = 
  \left( 
  v_{\theta}-(1-\eta) v_{\rm K} \right)^2 + 
  v_{\rm rad}^2 + 
  v_{\rm ver}^2 
  \,,
\end{align}
where $ v_{\rm theta},  v_{\rm rad},  v_{\rm ver}$ are respectively the azimuthal, radial and vertical velocity components of the embryo, and $(1-\eta)v_{\rm K}$ is the velocity of the gas.
By using this formulation of the relative velocity, we automatically take the eccentricity and inclination dependency of the pebble accretion rate into account.
Finally, the accretion radius is determined from the requirement that $t_{\rm fric} = t_{\rm def}$, which we do numerically by iteration.

\begin{figure}[h!]
  \centering
  \includegraphics[width=8.8cm]{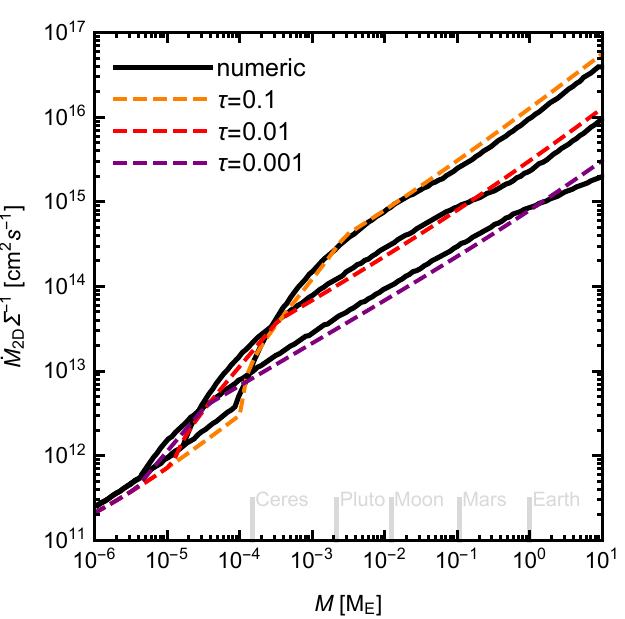} \\
  \caption{
  The product $2r_{\rm acc} v_{\rm acc}$, or equivalently, $\dot M_{\rm peb,2D}
/\Sigma_{\rm peb}$ as function of embryo mass.
  Dark lines represent the results of 2-body embryo--pebble integrations in 2D,
for particles with different Stokes number.
  The colored dashed lines correspond to the results from our accretion
prescription, covering strong coupling, the decrease to weak coupling, and
finally, around the smallest masses, the accretion purely following
gravitational deflection. 
  Values shown here are for an embryo at $2.5$\,AU with headwind velocity $\eta
v_{\rm K} = 7\times10^3$\,cm\,s$^{-1}$.
  The simulations presented in this paper start with embyos with a mass of
$0.01$\,M$_{\rm E}$. 
  }
  \label{fig:rv}
\end{figure}

In the above regimes, the crossing time past the embryo was always longer than the deflection time.
Around lower mass bodies, this may not longer be true. Therefore, in this so-called weak-coupling Bondi regime, we need the crossing time: 
\begin{align}
  t_{\rm cross} = 
  \frac{2 \sqrt{r_{\rm int}^2 - r_{\rm acc}^2 } }{v_{\rm acc}} \,.
\end{align}
Here, the nominator gives the length of the path of an unperturbed particle,
the chord, through the circle around the accretor with interaction radius
$r_{\rm int}$. 
In practice, we take $r_{\rm int}$ to be the smallest of the Bondi or Hill
radius $r_{\rm int} = \min(r_{\rm H}, GM_{\rm p}/v_{\rm rel}^2)$.
Therefore, when $t_{\rm cross} < t_{\rm def}$, we limit the accretion radius by requiring $t_{\rm fric} = t_{\rm cross}$. 
The latter expression can simply be solved analytically. 
We illustrate in Fig.\,\ref{fig:rv} the accretion prescription (dashed lines)
against 2D integrations performed in the shearing sheet, showing good agreement
including the transition to weak coupling, e.g. for $\tau_{\rm f}=0.1$-pebbles
from Pluto to Ceres mass, into the regime for a purely gravitational cross
section around even smaller masses.

\subsection{Pebble midplane}
\label{app:pa2}

In order to determine the accretion rate, we have to determine how much of the pebble flow falls within the accretion radius.
Pebbles settle towards the disc midplane \citep{Youdin_2007}. The pebble scale height is given by 
\begin{align}
  \frac{H_{\rm peb}}{H} \approx \sqrt{\frac{\alpha_{\rm z}}{\tau_{\rm f}}} \, 
\end{align}
where we have taken the vertical stirring parameter $\alpha_{\rm z}$ to be equal to the viscous $\alpha$.
This represents a well-settled particle layer in a nearly laminar midplane.

If the accretion radius starts exceeding the pebble scale height, we switch from 3D pebble accretion, 
\begin{align}
  \dot M_{\rm peb, 3D} = \pi r_{\rm acc}^2 v_{\rm acc} \frac{\Sigma_{\rm peb}}{\sqrt{2\pi} H_{\rm peb}} 
\end{align}
to a 2D accretion rate
\begin{align}
  \dot M_{\rm peb, 2D} = 2 r_{\rm acc} v_{\rm acc} \Sigma_{\rm peb}
\end{align}

Because of the low pebble scale height, it is possible for bodies with inclinations $i \gtrsim H_{\rm p}/r$ to escape from the pebble midplane and to stop accreting pebbles. 
We therefore simply cut the accretion rate $\dot M_{\rm peb} = 0 $, when the vertical position of the body exceeds the pebble scale height $z> H_{\rm peb}$.

\subsection{Pebble accretion in the 3D strong-coupling regime}
\label{ap:3dstrong}

We briefly derive the expression of Eq.\,(\ref{eq:3Dacc}).
Setting $t_{\rm fric}$ equal to $t_{\rm def}$, one obtains
\begin{align}
  r_{\rm acc}^2 v_{\rm acc} = t_{\rm fric} GM_{\rm p} \,.
\end{align}
In the 3D accretion regime the accretion rate becomes
\begin{align}
   \dot M_{\rm peb, 3D} 
   &\approx \pi r_{\rm acc}^2 v_{\rm acc} \frac{\Sigma_{\rm peb}}{\sqrt{2\pi} H_{\rm p}} \\
   &\approx \pi t_{\rm fric} GM_{\rm p} 
   \frac{F_{\rm peb}}{\sqrt{2\pi} H_{\rm p} (2\pi r) (2t_{\rm fric} \Omega_{K} \eta v_{\rm K})} \,,
\end{align}
where in the last line we expressed the surface density through the pebble flux $F_{\rm peb}$. 
Here, the product $2 t_{\rm fric} \Omega_{K} \eta v_{\rm K} $is the radial
drift speed of the pebbles. One finally obtains
\begin{align}
   \dot M_{\rm peb, 3D}
   &\approx \frac{1}{4 \sqrt{2\pi}} \frac{GM_{\rm p}}{r^2 \Omega_{\rm K}^2} 
   \frac{1}{\eta} 
   \frac{F_{\rm peb}}{H_{\rm p}} \\
   &= \frac{1}{4 \sqrt{2\pi}} \frac{M_{\rm p}}{M_\odot} 
    \left(\frac{H_{\rm p}}{H}\right)^{-1} 
    \left(\frac{H}{r} \right)^{-1} 
    \frac{1}{\eta}
    F_{\rm peb}
\,.
\end{align}
Therefore, when embryos are sufficiently large such that $t_{\rm def}>t_{\rm cross}$ and accretion proceeds in the 3D regime, growth is exponential in nature. 
Moreover, the expression has no orbital dependency in the viscously heated inner discs where we took the aspect ratio $H/r$ to be constant.
Because there is no longer an explicit dependency on the relative velocity the expression does not depend on the eccentricity, as long the strong coupling criterion is satisfied. 
This is not true in the 2D accretion regime \citep{Liu_2018}.
The accretion rate only depends on the particle size through the level of pebbles settling to the midplane $H_{\rm p}/H = \sqrt{\alpha_{\rm z}/\tau_{\rm f}}$.

Recently, \citet{Ormel_2018} numerically derived 3D accretion rates, which
differ from the 2D integrations discussed in Sec.\,\ref{app:pa1}. 
They find a similar scaling relation as Eq.\,(\ref{eq:3Dacc}), but measure accretion rates larger by approximately a factor $4$.
Therefore, following \citet{Ormel_2018}, we have moderately underestimated the filtering fraction in 3D for our choice of $\alpha_z$ and $\tau_{\rm f}$.
This implies similar growth rates as in this work could be obtained with, crudely, a factor $4$ smaller pebble flux or a factor $4$ larger pebble scale height. 
However, we do note that filtering efficiencies are also changed and higher
scale heights alter the accretion rates of inclined embryos, so this may lead
to differences.
Therefore, follow-up work will benefit from accurate prescriptions as in
\citet{Ormel_2018} and the future inclusion of the effect of the complex gas
flows around planetary bodies
\citep{Ormel_2013,Ormel_2015,LL_2017,Popovas_2018}.

\begin{figure}[h!]
  \centering
  \includegraphics[width=8.8cm]{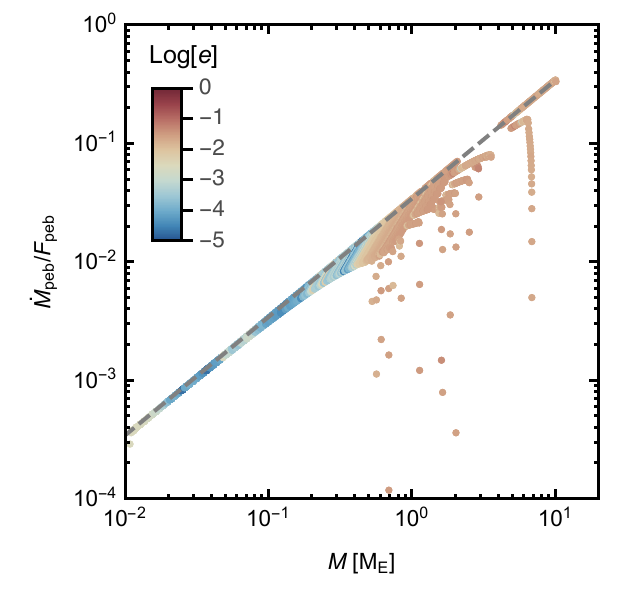} \\
  \caption{
  Accretion rate with respect to the total outer pebble flux, as function of embryo mass.
  The points show the embryos with their evolution shown in steps of $10^4$\,yr, for the $3$\,Myr duration of the gas disc phase (\texttt{runf5-1}).
  The colors correspond to the eccentricities of the bodies, 
  as indicated by the color bar.
The gray dashed line is the analytical expression Eq.\,(\ref{eq:3Dacc}), which represents 3D accretion in the strong coupling limit.
  The strong reduction in accretion rate at late time is mainly due to pebble
filtering by the outer embryos.
  }
  \label{fig:mf_run1}
\end{figure}

\subsection{Pebble accretion-driven drift}
\label{app:padrift}

Because of our prescribed accretion rates we do not automatically take into
account the angular momentum exchange between embryos and the population of
planetesimals and pebbles.
For embryos on the short orbits that we consider, type-I migration and damping dominates pebble-related effects. 
And, in this study, we choose to ignore the planetesimal population.
Nevertheless, we want to avoid the unphysical limit of accreting zero angular momentum material.
Therefore, we have implement a correction for pebble accretion-driven drift.
Consider the angular momentum balance as a body grows by a mass $dM$,
\begin{align}
  \sqrt{GM_{\odot}} (M+dM) (r+dr)^{1/2}  =  \sqrt{GM_{\odot}} M r^{1/2} + dM (v_\theta - v_{\rm hw}) r \,.
\end{align}
Here, the first term on the right hand side is the initial angular momentum and the second term is accreted angular momentum of the inwards drifting pebbles. For simplicity we will assume that the relative velocity is dominated by the headwind and ignore the dependency on the Stokes number.
This can be rewritten to give 
\begin{align}
  \frac{1}{v_{\rm hw}} \frac{dv_\theta}{dt} = \frac{1}{M}\frac{dM}{dt} \,. 
\end{align}
This drag force would result in inwards drift. 
In our code, we have balanced this force against the artificial drift driven by adding mass without angular momentum transport. 
In this way we find an expression for a correction force along the azimuthal direction of the form
\begin{align}
  \frac{d v_\theta}{dt} = \frac{(1-\eta)v_{\rm K}}{M} \frac{dM}{dt}\,.
\end{align}
Note we have made several assumptions here (low $e$,$i$, no angular momentum transfer to spin, surrounding gas or a pebble accretion disc). However, as mentioned, this procedure is mainly here to avoid the unphysical limit of mass growth without any angular momentum exchange.

\end{document}